\documentclass[
reprint,
 amsmath,amssymb,
 aps,
floatfix,
]{revtex4-2}

\usepackage{graphicx}
\usepackage{dcolumn}
\usepackage{bm}

\usepackage{xcolor}

\begin{document}

\title{Theoretical foundations of nonlinear electromagneto-mechanical systems:\\ applying Finsler geometry to Kibble Balances}

\author{Pierluigi Rizza}
\email{pierluigi.rizza@polito.it}
\affiliation{INRIM, Istituto Nazionale di Ricerca Metrologica} 
\author{Patrizio Ansalone}%
\email{p.ansalone@inrim.it}
\affiliation{INRIM, Istituto Nazionale di Ricerca Metrologica}


\date{\today}

\begin{abstract}
Kibble balances are energy-conversion devices currently employed to realize the metrological redefinition of the kilogram unit of mass. The authors, noticing that the literature lacks a fundamental mathematical description of the Kibble balance working principle, propose a differential geometric approach to study general non-linear electromagneto-mechanical systems. Based on the application of Finsler geometry, the mathematical apparatus introduced in the paper is used to analyze the limits of modern Kibble balances, with particular attention to the magnetic design. The authors review the formalism introduced by G. Kron (who proposed a unified theory of electrical machines) and extend it to non-linear systems. Numerical simulations are presented to show the difference between a linear and non-linear analysis of Kibble balance machines, with attention to the hysteretic behaviour of the permanent magnet system.
\end{abstract}

\maketitle


\section{Introduction}

Electro-mechanical machines are devices based on the conversion of different forms of energy, namely mechanical and electromagnetic energy. In the early years of this new science, electro-mechanical machines were built following a \emph{ trial-and-error} approach, caused both by the absence of a  solid theoretical background and  the great difficulty in applying the new theoretical concepts to actual real-life scenarios. Nowadays, despite the development of both theoretical and numerical methods and the predominant use of computer simulations, the same time/money-consuming trial-and-error approach is still used and scientists have to constantly face the problem of introducing \emph{ad hoc corrections} to pre-existing and consolidated  electro-mechanical models each time they discover some discrepancy between theory and experimental results. While this is standard routine for all branches of experimental science, the fields involved with electromagnetism seem the most problematic. Indeed, scientists cannot blindly adapt electromagnetic models to specific scenarios, since any minor change in the model risks to alter the already delicate eco-system of electromagnetic equations: new corrections, if not carefully considered, may inconsistently alter energy and field equations and may ultimately lead to un-reproducible results. 
Now more than ever, the problem is crucial: due to the SI redefinition of 2019, the kilogram is defined based on universal constants, which represent the building blocks of all units of measure \cite{mills2005redefinition, schlamminger2018redefining}. In particular, two promising primary methods to realize the kilogram exist: one of these methods aims at defining the kilogram using electrical units. It employs an electro-mechanical machine that uses voltage and resistance standards,  which heavily rely on the theory of the Josephson and Quantum Hall effects \cite{mills2005redefinition, milton2010quantum}. These machines are grouped under the name of \emph{Kibble balance} systems, in honour of the scientist Bryan Kibble who first introduced the device \cite{kibble1976measurement}. Initially, the machine was built to give more accurate values of the Planck constant. However, with the SI redefinition, the numerical value of the Planck constant (among others) was set as a fundamental constant without any associated uncertainty. Since then, the Kibble balance working principle has been inverted to define the kilogram unit. \\
The Kibble balance principle is (apparently) straightforward and is based on the energy conversion between electromagnetic and mechanical power: a current flowing in a coil immersed in a magnetic field is subjected to a force, and such a force is compared to the weight generated by a mass; thus the Newton and the Kilogram are metrologically realized. It can be argued that, indeed, the most fundamental quantity here is the force rather than the mass since the evaluation of local gravitational acceleration is required to obtain a mass value (with more associated uncertainty).\\
Kibble balances are based on a two-phase measurement: the velocity mode and the weighing mode. In the weighing mode, a current flows in a coil, where a pan is attached and a mass is put on top: the pan is balanced thanks to the current flowing in the coil. In the velocity mode, instead, no current flows in the coil and the mass is removed. The coil is, then, moved at a controlled constant speed inside the magnetic field. The necessity of the two-phase measurements relies on the difficulty of evaluating the infamous $Bl$ factor, which is associated with non negligible measurement uncertainties. Therefore, by equating quantities measured in both velocity and weighing mode, we avoid measuring the $Bl$ factor and we obtain the weight (and therefore the mass) of the mass under test. The equivalence between the system in the weighing and in velocity mode might be regarded as either true or false depending on the uncertainty required in the experiment. At extremely low uncertainties, this equivalence may not  be true anymore \cite{li2017coil}: for instance, the permanent magnet in the \emph{velocity-mode-system} and the permanent magnet in the \emph{weighing-mode-system} may as well be regarded as two distinct magnets of their own. Understandably, in the case of the Kibble balance, very low uncertainties are needed to establish consensus values to properly realize a generally accepted redefinition of the kilogram. Nowadays, metrologists spend great effort to reduce the various uncertainties contributions, especially the ones arising from the non-linear behaviour of the system, both related to magnetic hysteresis \cite{li2017coil, li2019investigation} and to mechanical hysteresis \cite{keck2022design}.
In these cases, ad hoc mathematical tools and corrections are developed each time (e.g. \cite{li2017coil}) to account for all the non-linearities involved in that particular system. This approach is often circumstantial to the specific experimental set-up or problem and might be modified again once a new set of corrections have to be made. \\
For these reasons, the authors believe that a fortiori mathematical foundation of non-linear electro-mechanical systems able to generalize the Kibble-Balance system is needed. The formalism should be able to consistently account for the non-linear behavior of the system to systematically control it: through the development and refinement of such a theory, more efficient (or even cheaper) versions of the Kibble Balance can be built.\\ The mathematical foundation here proposed is based on the theoretical approach introduced by Gabriel Kron in 1934 \cite{kron1934non} and is nowadays often adopted in disguise, without realizing the profound mathematical structure laying underneath. The spirit behind such a formalism was due to the recent theory of General Relativity, at that time, and the new interests of mathematicians and physicists in differential geometry \cite{lynn1958tensor, gibbs1952tensors, gibbs1965foundation}. Kron was deeply fascinated by the level of generalization conveyed by differential geometry and tensor calculus, which were not as popular as they may be now.  However, formalism has been applied only to linear machines without considering any hysteresis phenomena. These problems, in addition to more and more computer codes for FEM simulations being used in every engineering field, led to the abandonment of the theoretical work of Kron and his few successors \cite{von1968new}. \\
This paper will review the concept behind Kron formalism and extend it to general non-linear systems. As in Kron's work, we will introduce a formalism which revolves around the concept of the \emph{metrics}, which will be the instrument used to include the non-linear behaviour of the system, regardless of its mechanical or electro-magnetic nature. The method will be applied to a torque-generating machine that employs the Kibble balance principle. A torque Kibble Balance machine (as realized in \cite{hamaji2021design}) is preferred to a usual axial Kibble balance machine so that the reader can compare the calculations with the original work presented by Kron \citep{kron1934non} was based on rotating machines. 

\section{Electromechanical metric tensor}
 According to \cite{kron1934non}, the state of an electric machine, modelled as a lumped network, can be identified with a tangent vector to a point in a general non-Riemannian $N$-dimensional manifold; the basic idea is that the equations of motion, governing the evolution of the state of the electrical network, can be written in the same form as the usual classical mechanics equations. For instance, comparing Newton's second law for a single body  to a circuit with an inductor, namely:
 \begin{align}
F=m\frac{d v}{dt}, \qquad \mathrm{e.m.f.}=L\frac{di}{dt}
 \end{align}
we can identify the inductance $L$ with an \emph{electrical inertia}, in the same way \emph{mass} is regarded as a mechanical inertial mass $m$. In the same manner, the force is identified with the  \emph{electromotive force}: in a simplistic hydrodynamic analogy, the current which \emph{flows} through each r-edge of a lumped electrical network $i^r$ is defined as the first time derivative of the charge $q^r$ flowing in the same edge; similarly, the velocity is the first time derivative of the position vector $\boldsymbol{x}=(x^1, x^2, \dots,x^k, \dots, x^n)$:
\begin{align}
i^r=\frac{dq^r}{dt}, \qquad v^k=\frac{dx^k}{dt}
\end{align}
If there are $N_E$ electrical elements (i.e., a capacitor, coils, a solenoid, \dots), and $N_M$ mechanical elements, then we can distinguish between electrical and mechanical  coordinates, both collected in a vector of generalized coordinates labeled as $z^{a}$. The mechanical coordinates are indicated by $x^k$ while the electrical ones are denoted by $q^{r}$. Thus we can define a space of $N=N_E+N_M$ dimension. In the following, the generalized velocity $\dot{z}^{a}$ denotes the actual velocity $v^k$ and the currents $i^r$.\\
Kron's formalism aims to identify \emph{invariant quantities} to be used as system constants.   Possible candidate invariants of a system are scalar quantities: a clear example is the principle of  conservation of energy (indeed a scalar!). The question is: how to obtain invariant scalars from vectors and, in general, tensor quantities? The procedure is quite common in simple mechanical systems. \\
Recall that the mechanical kinetic energy is defined as $T=\frac{1}{2}mv^2$.
In analytical mechanics \citep{abraham2008foundations}, a manifold is introduced to describe the mechanical system. In a manifold, a metric tensor $g_{uv}$ is introduced as a way of determining distances $dl$, that is:
\begin{align}
\label{eq:line_elem}
dl=\sqrt{g_{uv}(x)dx^u dx^v}
\end{align}
where the Einstein summation convention for repeated indexes is assumed. As an example, using Cartesian coordinates, the metric tensor is the identity matrix and, indeed, along the axis $e^u$, the distance between two points is simply the distance $dl=dx^{u}$; however, describing the same phenomenon in cylindrical or spherical coordinates will employ different metric tensors, but still give the same distance! 
From analytical mechanics, speed is evaluated as $v^2 \equiv g_{ab}\frac{dx^b}{dt}\frac{dx^a}{dt}$ and the kinetic energy is simply:
\begin{align}\label{eq:kin_similar}
T=\frac{1}{2}g_{ab}v^av^b
\end{align}
where $g_{ab}$ are the coefficients of the inertia tensor. This tensor contains the information on the masses involved: for instance, given $n$ particles with masses $m_1$, $m_2$, $\dots$, then $g_{ab}=\delta_{ac}m_{cb}$ ($\delta_{ab}$ is the identity matrix) and the kinetic energy is the usual $T=1/2m_1(v^1)^2+1/2m_2(v^2)^2+\dots$. We see that the kinetic energy is, by construction, a quadratic form in the velocity. This is not true for general non-linear systems, such as electromagnetic systems.

\subsection{Electromagnetic energy}
In the case of linear electromagnetic systems, we can define magnetic energy as:
\begin{align}
T_H=\int_{\Omega_V}dV \int d\mathbf {B}\cdot\mathbf {H}=\frac{1}{2}\int_{\Omega_V} dV\  \mathbf {H}\cdot\mathbf {B}
\end{align}
where $\mathbf{B}$ and $\mathbf{H}$ are (respectively) the magnetic flux density vector and the magnetic field.
Considering the integral form of Maxwell equation ($\mathbf{J}$ denotes the current density vector):
\begin{align}
\int_{S^a}\mathrm{rot}\mathbf {H}d\mathbf {S}=\int_{S^a}\mathbf {J}d\mathbf {S}=i^a
\end{align}
and using Stokes theorem:
\begin{align}
\oint_{\partial S^a} \mathbf {H}d\mathbf {l}=i^a,
\end{align}
it is shown how the circuital current $i$ along a closed path can be considered equivalent to the circuitation of the $\mathbf {H}$-field. 
The assumption of the equivalence between the field representation and the lumped parameter model is the theoretical basis of Kron's analysis: the $\mathbf {H}$ field is taken to be equal to a current per unit length, where this length is the characteristic length for the given electromagnetic element (e.g. length of a coil). If the coil has $N$ turns, then the equivalent current is $N\cdot i$, where $i$ is the current associated with a single-turn coil.\\
Bearing this in mind, the field $\mathbf {H}$  generated by different current sources $i^{a}$ (which can include magnetizing currents when permanent magnets are employed) is represented by $\sum \mathbf {w}_a i^a$ with $\mathbf{w}_a$ are the Whitney basis \cite{bossavit1988whitney} this representation means that, if we construct a space where the basis are the currents $i^{a}$, then the total magnetic field $\mathbf {H}$ can be written in terms of its basis as $\sum \mathbf {w}_a i^a$. In general, the $\mathbf{B}$ and $\mathbf{H}$ fields can be thought as linked by the magnetic susceptibility tensor, which can be reduced to a scalar in the case of linear electromagnetic systems. In the linear case, denoting by $\mu$ the magnetic susceptibility, it follows that the total magnetic energy can be written as 
\begin{align}
T_H=\frac{1}{2}\int_{\Omega_V} dV \mu \ i^a\mathbf {w}_a\cdot\mathbf {w}_b i^b
\end{align}
We can rewrite the above equation, introducing the 
magnetic inductance tensor $L_{ab}$:
\begin{align}
T_H=:\frac{1}{2}L_{ab}i^ai^b
\end{align}
and invite the reader to
note the similarity with the form of \eqref{eq:kin_similar}. \\
We can construct a \emph{multi-physics} tensor through the direct sum of the inertia tensor and the inductance tensor, to account for both mechanical and electrical aspects of the systems; this tensor is what we call the \emph{electro-mechanical metric tensor}:
\begin{align}
G_{\alpha\beta}=L_{\alpha\beta}\oplus g_{\alpha\beta}
\end{align}
where the operator $\oplus$ means in this case:
\begin{align}
\mathbf {G}=\begin{bmatrix}
\mathbf {L}&\mathbf {0}\\
\mathbf {0} & \mathbf {g}
\end{bmatrix}
\end{align}
If we now consider the generalized coordinates $z^{a}$, constructed from the mechanical coordinates $x^{a}$ and electrical coordinates $q^{a}$ (as introduced at the beginning of this paragraph), the total energy (which is a scalar) can be written as:
\begin{align}
T=\frac{1}{2}G_{a b}\dot{z}^{a}\dot{z}^b\qquad a,b=1,2, \dots N_E, \dots N_M
\end{align}
The kinetic energy is once again defined using a metric tensor and the first time derivative of the generalized coordinates (in the same way as \eqref{eq:kin_similar}).
\subsection{Metric in Kron's space}
In the case of non-linear permanent magnets (PMs), the situation appears more complex. In this case, the magnetic energy of the system can be written as:
\begin{align}
\label{eq: energy_kron}
w_{em}(x^{a}, i^{a})=w^\star_{em}+\int_{\Omega_V}dV \int d\mathbf {B}\cdot\mathbf {H}
\end{align}
where $w^\star_{em}$ is the self energy of the PM and $\Omega_V$ is referred to the domain of the PM. 
Starting from the idea introduced in \cite{kondo1955raag}, we postulate a form of  constitutive equation of the $\mathbf{B}$-field depending on the $\mathbf{H}$-field and on the magnetization field $\mathbf{M}$, namely:
\begin{align}
\mathbf {B}(\mathbf {H}, \mathbf {M})=:\mathcal{F}[\mathbf {H}(x^{i}, i^{a}), \mathbf {M}(x^{i}, i^{a})]
\end{align}
where $\mathcal{F}$ is a non-linear vector function. 
Computing $d\mathcal{F}$, the integral becomes:
\begin{eqnarray} \label{eq: non_lin_constit}
\int d\mathbf {B}\cdot\mathbf {H}=\int dx^{i}\bigg(\frac{\partial \mathcal{F}}{\partial \mathbf {H}}\frac{\partial \mathbf {H}}{\partial x^{i}}+\frac{\partial \mathcal{F}}{\partial \mathbf {M}}\frac{\partial \mathbf {M}}{\partial x^{i}}\bigg)\cdot\mathbf {H}+\nonumber\\
+\int di^{a}\bigg(\frac{\partial \mathcal{F}}{\partial \mathbf {H}}\frac{\partial \mathbf {H}}{\partial i^a}+\frac{\partial \mathcal{F}}{\partial \mathbf {M}}\frac{\partial \mathbf {M}}{\partial i^{a}}\bigg)\cdot\mathbf {H}
\end{eqnarray}
If we focus on the case of  magnetic elements fixed in space (no relative movement), we can neglect the first term to define a non-linear magnetic flux linkage $\varphi_a$ associated to the $i$-th current $i^a$, which is related to how the magnetic energy changes with respect to the currents:
\begin{align}
\label{eq: flux_link}
\varphi_a=:\int_{\Omega_V} dV  \bigg(\frac{\partial \mathcal{F}}{\partial \mathbf {H}}\frac{\partial \mathbf {H}}{\partial i^a}+\frac{\partial \mathcal{F}}{\partial \mathbf {M}}\frac{\partial \mathbf {M}}{\partial i^a}\bigg)\cdot\mathbf {H}
\end{align}
Finally, the magnetic energy \eqref{eq: energy_kron} can be written as:
\begin{align}
w_{em}=w_{em}^\star+\int di^a\varphi_a\label{eq:eq_linear}
\end{align}
As in classical mechanics, where one defines the linear momentum $p_i$ in terms of the velocities through a constitutive equation of the type $p_{i}=g_{ij}v^j$, we introduce the following constitutive equation between fluxes and currents:
\begin{align}
\varphi_a=L_{ab}i^{b}
\end{align}
The new generalized metric $L_{ab}$ represents the bridge between two dual quantities.\\
Since introducing dual quantities is related to the introduction of a metric tensor \cite{abraham2008foundations}, one can view the non-linear behavior as the occurrence of the new (Kron's) metric tensor $L_{ab}$, representing the coupling between electrical elements. In particular, if the metric tensor is symmetric, it can be diagonalized, and the \emph{anisotropic} behavior (in this new generalized sense) can be overcome in a proper coordinate system. 
Finally, from the derivation of the non-linear magnetic energy, the metric shows non-Riemannian characteristics since it also depends on the generalized velocities (in this case, just the currents):
\begin{align}
L_{ab}=L_{ab}(x^{a}, i^{a})
\end{align}
These considerations can be also applied to systems affected by mechanical hysteresis (as often happens in systems with hinges with back-and-forth motion): in the usual mechanical constitutive equation $p_i=g_{ij}v^j$, the mechanical metric tensor may be regarded as velocity dependent $g_{ij}=g_{ij}(x^\alpha, \dot{x}^\alpha$). To account for these dependencies, one can easily introduce a generalized metric tensor of the form:
\begin{align}
L_{ab}=L_{ab}(z^{a}, \dot{z}^{a})
\end{align}
where the metric tensor is dependent on the generalized higher order terms, namely currents and velocities.\\
Now we are ready to derive the equations of motion, which can be obtained following the standard and consolidated procedures of classical mechanics \cite{abraham2008foundations}: once we have defined a suitable generalized metric, we assume a general representation of the total energy of the system, to which it is possible to apply the usual variational principles.   If the metric is independent of higher-order terms ($\partial L_{ab}/\partial\dot{z}^a=0$), the usual minimization principles can be easily applied and the related Christoffel symbols can be obtained from the derivative of the metric with respect to the generalized coordinates $z^\alpha$ (see Appendix \ref{par:riem}). However, with the introduction of a metric dependent on higher order terms, the magnetic and kinetic energy cannot be brought directly into a quadratic form. Therefore, a new \emph{geometry} must be introduced: Finsler geometry. Apart from its deep mathematical structure, such a geometry will only be used to write easier \emph{linear-look-alike} equations of the non-linear system, which can be then solved numerically. To this aim, we present a brief derivation of the governing equation arising from the tools of Finsler geometry.

\section{Finsler metric}
\label{par:finsler}
When analyzing electro-mechanical systems, we pointed out two cases: in the first case, the metric is independent of the velocity/currents. In this case, the magnetic and kinetic energy is a quadratic form in the generalized velocities and the usual equations of motion are derived as in \cite{kron1934non}. In the second case, the metric depends on the velocity/currents. The magnetic and kinetic energy cannot be brought directly into a quadratic form if the usual Riemannian metric is employed. To overcome this problem, we introduce a new metric, the \emph{Finsler metric}  (the reader is also referred to \cite{ratliff2021generalized, boccaletti1997finsler} for some practical applications in robotics and cosmology). In the case of non-linear and hysteretic phenomena, we saw that the total energy of the system \eqref{eq: energy_kron} can be written as:
\begin{align}
w(z^{i}, \dot{z}^{i})=\int_{0}^{\dot{z}^{a}}d\dot{z}^{a}\varphi_{a}+\int_{0}^{\dot{z}^{u}}d\dot{z}^{u}p_u
\end{align} 
where $\varphi_a$ is the non-linear magnetic flux linkage and $p_u$ is the non-linear mechanical momentum. The term involving $p_u$ can be derived from \eqref{eq: non_lin_constit} in the same way as done for \eqref{eq: flux_link}, not neglecting the first integral term.
The state space origin of the system is shifted to the initial generalized velocities (i.e. $\dot{z}^{\beta}_0=0$).
In order to handle the non-linear characteristics of the system, we follow the idea presented in \cite{kondo1955raag, kawaguchi1977application} and introduce the Finsler metric $\gamma_{ik}$ by defining the invariant line element $dl$:
\begin{align}
dl^2=\gamma_{jk}(z, \dot{z})dz^jdz^k
\end{align}
Note that the above equation is similar to \eqref{eq:line_elem}, apart from the dependence on the higher order term $\dot{z}$.
If we introduce time $t$ as an additional parameter, we study the problem in a $(N+1)$ dimensional manifold where $z^{0}=t$.\\
We now sketch the derivation of the Finsler metric (the reader is referred to \cite{kawaguchi1977application} for a detailed explanation). We start by introducing a new coordinate $\xi$ (which will be identified as time) to write the energy into a quadratic form. We introduce a new energy representation $\tilde{w}$
\begin{align}
\tilde{w}=\tilde{w}(z^{\alpha}, \xi^{\beta}), \qquad\alpha=0,1,\dots
\end{align} where the relation between the new generalized velocities $\xi^{\alpha}$ and the old velocities $\dot{z}^{a}$ is:
\begin{align}
\xi^{\alpha}=\xi^{0}\dot{z}^{a} \qquad a=1,2,\dots
\end{align}
Since the two energies must be the same, apart from their representation, we impose the following:
\begin{align}
\tilde{w}(z^{\alpha},\xi^{\beta})=(\xi^{0})^2 w(z^{a}, \dot{z}^b)
\end{align}
From this, we define the new metric as:
\begin{align}
\gamma_{\alpha\beta}=\frac{\partial^2\tilde{w}}{\partial \xi^{\alpha}\xi^{\beta}}
\end{align}
By a lengthy calculation, explained in detail in \citep{kawaguchi1977application},
the non-null terms of the Finsler metric are: $ \gamma_{ab}=\frac{\partial^2 w }{\partial \dot{z}^a\partial \dot{z}^b}$
which represent the new self/mutual-inductance terms; $
\gamma_{00}=2(w- \varphi_{a}\dot{z}^{a}-p_{u}\dot{z}^{u}+\frac{1}{2} L_{ab}\dot{z}^{b}\dot{z}^{a}
+\frac{1}{2} g_{uv}\dot{z}^{u}\dot{z}^{v})$ whose meaning is related to the difference between energy and co-energy, which indeed is zero for linear systems;
$\gamma_{0a}=\varphi_a-L_{ab}i^{b}$ for electrical parts and
$\gamma_{0a}=p_u-g_{uv}v^v$ for mechanical parts, which are both zero for linear systems. \\
From the above equations, it can be seen that introducing higher order terms in electromagnetic metric terms $L_{ab}$ and in the mechanical metric terms $g_{uv}$ can lead to a more comprehensive description of both the mechanical and electromagnetic non-linearities. However, we focus this paper only on the magnetic non-linearity, therefore assuming $\partial g_{uv}/\partial z^{v}=0$.
\\
Introducing external forces and a Rayleigh dissipation function $\mathcal{R}$, the EL-equations are written as:
\begin{align}
\frac{d}{dt}\frac{\partial w}{\partial \dot{z}^{\alpha}}-\frac{\partial w}{\partial z^{\alpha}}+\frac{\partial V}{\partial z^\alpha}+\frac{\partial \mathcal{R}}{\partial \dot{z}^{\alpha}}=0
\end{align}
where $z^0=t$.
The equations of motion are, therefore:
\begin{align}
\gamma_{\alpha\beta}\frac{d^2z^\beta}{dt^2}+\Gamma_{\beta \eta\alpha}\frac{dz^\beta}{dt}\frac{dz^\eta}{dt}+r_{\alpha\beta}\frac{dz^\beta}{dt}=f_\alpha
\end{align}
where we have introduced the Christoffel symbols $\Gamma_{\beta \eta\alpha}$:
\begin{align}
\Gamma_{kji}=\frac{1}{2}\bigg(\frac{\partial \gamma_{ji}}{\partial z^{k}}+\frac{\partial \gamma_{ki}}{\partial z^{j}}-\frac{\partial \gamma_{kj}}{\partial z^{i}}\bigg)
\end{align}
where, again, it can be seen that no higher order derivative $\partial\dot{z}^j$ is included, exactly like for the linear case (except that the Finsler metric already takes care of the non-linear dependencies).
Denoting $e_{a}$ the voltage, $r_{ab}$ the resistances, $f_{u}$ the force (or torque, depending on hoe the $\gamma_{ij}$ tensor is defined), the governing equations are explicitly written as:
\begin{eqnarray}
e_a= \gamma_{ab}\frac{di^{b}}{dt}+\dot{x}^u\partial_u \gamma_{0a}+i^b\dot{x}^u\partial_u\gamma_{ab}+r_{ab}i^b
\end{eqnarray}
\begin{eqnarray}
f_u=\gamma_{uv}\ddot{x}^v-\frac{1}{2}i^ai^b\partial_u\gamma_{ab}+\nonumber\\
-i^a\partial_u\gamma_{a0}-\frac{1}{2}\partial_u \gamma_{00}+r_{uv}\dot{x}^v
\end{eqnarray}
In the above equations, the introduction of the dissipation function yields the electrical resistance and mechanical friction terms.
From the above equations, one can see the usual intuitive characteristics of the Kibble equations, mixed with the usual electrical and Newtonian equations. 
The voltage $e_a$ is dependent on the relative motion of the components ($\dot{x}^u\partial_u \gamma_{0a}$) and the variation of all the currents involved in the systems ($\gamma_{ab}\frac{di^{b}}{dt}$); currents and motion are intrinsically intertwined ($i^b\dot{x}^u\partial_u\gamma_{ab}$). In the same way, the force depends on the acceleration of the electrical elements ($\gamma_{uv}\ddot{x}^v$, which is simply $F= ma$) and on the relation of all the currents involved ($\frac{1}{2}i^ai^b\partial_u\gamma_{ab}+i^a\partial_u\gamma_{a0}$) which generalize the Kibble equation to non-linear interactions between PM and coils; finally, additional mechanical and electrical hysteresis are inserted in $\gamma_{00}$ terms. Thus, we proved that to describe the non-linear behavior of the systems, scientists can start by considering the proposed model and conveniently discarding some non influential non-linear terms, rather than adding new correction starting from the linear model and having to worry for any fundamental inconsistency.

\subsection{Verification of Finsler metric coefficients}
We recall that the energy of the system is, in anyway, the same regardless of its representation. For instance, consider the case of a system comprised of only two non-linear electromagnetic elements (e.g. coils) denoted by indexes $\alpha,\beta$. Consider $\dot{z}^{\alpha}=i^{\alpha}$ and $\dot{z}^{\beta}=i^{\beta}$. Using the Finsler metric, the magnetic energy can be written in the desired quadratic form:
\begin{eqnarray}
\tilde{w}=\frac{1}{2}\gamma_{\alpha\beta}i^{\alpha}i^{\beta}=\frac{1}{2}\gamma_{00}+
\gamma_{0a}i^a+\nonumber\\+\gamma_{0b}i^b+\frac{1}{2}\gamma_{aa}i^{a}i^{a}+\frac{1}{2}\gamma_{bb}i^{b}i^{b}+\gamma_{ab}i^{a}i^b
\end{eqnarray}
In this case, $\tilde{w}$ and $w$ from \eqref{eq:eq_linear} are two representation of the same physical quantity.
From \eqref{eq:eq_linear}, we deduce that for our two-element system, each representation must satisfy the following conditions \citep{von1968new}:
\begin{align}
\begin{split}
\varphi_{a}=\frac{\partial w}{\partial i^{a}}, \qquad \gamma_{aa}=\frac{\partial^2 \tilde{w}}{\partial i^{a}\partial i^{a}}\\
\varphi_{b}=\frac{\partial w}{\partial i^{b}}, \qquad \gamma_{bb}=\frac{\partial^2 \tilde{w}}{\partial i^{b}\partial i^{b}}\\
\gamma_{ab}=\frac{\partial^2 w}{\partial i^{a}\partial i^{b}}=\frac{\partial^2 w}{\partial i^{b}\partial i^{a}}
\end{split}
\end{align}
From these conditions, we get the constraints:
\begin{eqnarray}
\frac{\partial \gamma_{00}}{\partial i^{c}}=\frac{\partial \gamma_{ab}}{\partial i^c}i^ai^b, \nonumber\\ \frac{\partial \gamma_{0a}}{\partial i^{c}}=-\frac{\partial \gamma_{ab}}{\partial i^c}i^b, \nonumber\\ \frac{\partial \gamma_{0b}}{\partial i^{c}}=-\frac{\partial \gamma_{ab}}{\partial i^c}i^a
\end{eqnarray}
Since the self and mutual inductance terms can be experimentally derived, the other terms of the metric can be obtained by using the previous constraints, apart from current independent terms.

\section{Numerical simulations}
\begin{figure}
\centering
\includegraphics[width=8.5cm]{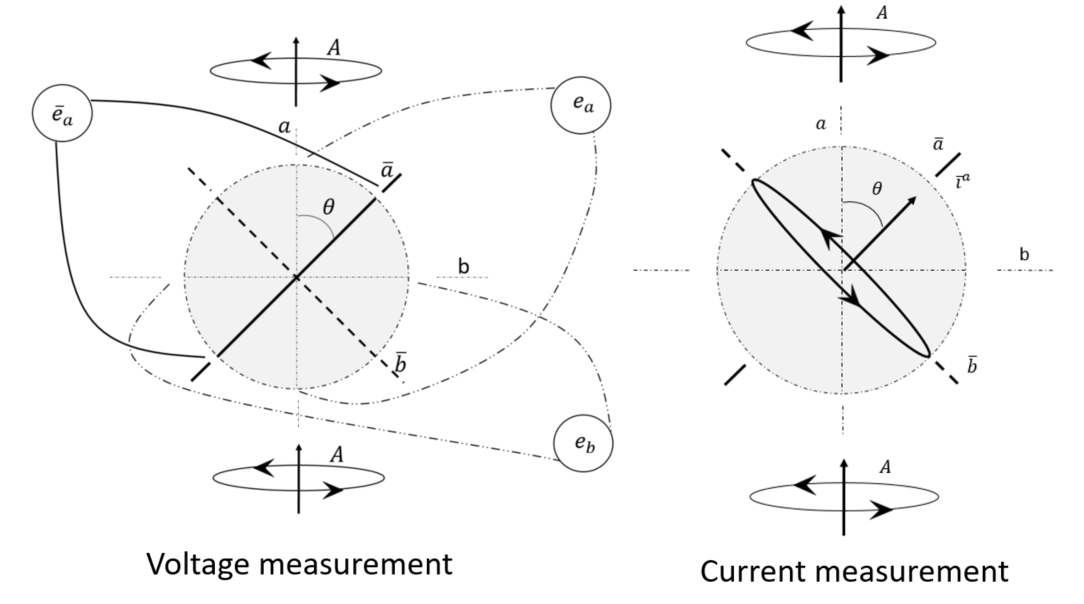}
\caption{Convention of axis orientation: a current flowing in a coil is identified and denoted with the coil axis, which corresponds to the direction of the magnetic field induced by the same current ($\bar{a}$-axis in the right part of the picture). The voltages, on the other hand, are thought to lie in the coil plane (left part of the picture).}\label{fig:axis_ref}
\end{figure}

 In this section, we will show a simple application on a torque-generating machine, employing the principle of the Kibble balance, using the two-phase measurement principle \cite{li2022irony}.  The choice of a torque machine (see \cite{hamaji2021design}) is preferred to an axial one so that the reader can compare the obtained results with theoretical ones existing in the literature \cite{kron1934non, von1968new}.
We apply the previous formalism to simulate what happens during the weighing mode in case of current reversal (highly exaggerating what happens in a mass-on/mass-off measurement \cite{li2017coil}) to see if the measured force exerted by the coil suffers from any non-linear behavior. In practice, we show the simulation of a \emph{worst case scenario}, where highly non-linear properties can manifest to understand how the mathematical formalism introduced can be easily employed, even in such cases.  \\ 
We present the case where a coil can rotate in a magnetic field perpendicular to the coil axis. We start by defining the generalized metric from the self-inductance and mutual inductance terms. To do this, we must establish proper reference systems: this is essential, since the metric quantities vary according, for instance, to their relative position (e.g. mutual inductance).
The axis convention adopted is the one often used in the theoretical analysis of electrical machines \cite{kondo1955raag}: capital Latin letters are used for the stator axis; lower Latin letters are used to denote the fixed frame on the rotor, while over-lined lower Latin letters are used for the rotor co-moving axis; finally when no distinction between the frame is to be made, Greek letters are used.
The setup is schematically depicted in figure \ref{fig:axis_ref}: the magnets on the stator are represented by their equivalent coil counterpart and generate a field along the $A$-axis; the rotor consists of a coil whose axis is along the $\bar{a}$-axis.
In the co-moving frame of the rotor, we identify the orthogonal axes $a,b$. 
In the stationary frame where axes of both rotor and stator are fixed, the inductance matrix has the following representation:

\begin{align}
[L_{\alpha\beta}]=
\begin{bmatrix}
L_{AA} & M & 0 \\
M & L_{aa} & 0  \\
0 & 0 & L_{bb}\\
\end{bmatrix} 
\end{align}

where the inductance terms are non-linear with respect to the currents. $L_{AA}$ is the self-inductance term of the PM, $L_{aa}$ and $L_{bb}$ are the self-inductance terms for the rotor coil in the co-moving frame,  and $M$ is the mutual inductance between the rotor coil and the PM in the stator. Even though there is just one coil on the rotor, we still define the metric considering both quadrature axis $L_{aa}$ and $L_{bb}$: this is because Kron's metric is also able to show how the force and voltage measurements change according on \emph{where} such measurements are performed
\subsection{Building the Finsler metric}
In this section, we apply the above concepts to a highly non-linear system. As suggested in \cite{von1968new}, the non-linear self-inductance of the electromagnetic elements can be expressed as an exponentially decaying function of its current $i^{k}$ in the form:
\begin{align}
L_{kk}(i^k)=M\exp(-A|i^{k}|)
\end{align}
This relation represents the non-linear self-inductance corresponding to a specific non-linear B-H curve, which has to be experimentally characterized.\\
We assume the following  generic representation of the total magnetic energy of the system:
\begin{align}
w=\alpha(i^{A})^2+\beta(i^{b})^2+\xi i^{A}i^{b}+\eta (i^A)^2i^b
\end{align}
In this case, it is not immediately possible to write the energy starting from the usual current independent generalized metric $G_{ab}$. If this were possible, we would have:
\begin{align}
w=\frac{1}{2}L_{AA}(i^{A})^2+\frac{1}{2}L_{bb}(i^{b})^2+L_{Ab}i^{A}i^b
\end{align} 
which is true only if $\eta=0$; in this case, we would have  $L_{aa}=\alpha$, $L_{bb}=\beta$ and $L_{ab}=\xi$ as usual, and the problem collapses to a linear one. However, if the condition $\eta\neq 0$ arises from experimental data, the only possibility is that the mutual inductance $L_{Ab}$ is current dependent, therefore the generalized metric tensor $G_{ij}$ is current dependent.\\
Our aim is, then, to write the energy using the Finsler metric to show how this approach automatically takes care of the nonlinearity of the problem. In terms of the Finsler metric, the energy is:
\begin{eqnarray}
w=\frac{1}{2}\gamma_{AA}(i^{A})^2+\frac{1}{2}\gamma_{bb}(i^{b})^2+\gamma_{Ab}i^{A}i^b+\nonumber\\+\gamma_{0A}i^A+\gamma_{0b}i^{b}+\frac{1}{2}\gamma_{00}
\end{eqnarray}
From Section \ref{par:finsler}, the Finsler metric is:
\begin{align}
\gamma_{\alpha\beta}=\begin{bmatrix}
2\eta (i^A)^2i^b & -2\eta i^A i^b & -\eta (i^A)^2\\
-2\eta i^A i^b & 2(\alpha+\eta i^b)& \xi + 2\eta i^A\\
-\eta (i^A)^2 &\xi + 2\eta i^A & 2\beta
\end{bmatrix}
\end{align}
It can be seen that when $\eta=0$, $g_{00}=g_{0a}=g_{0b}=0$ and we recover the current independent metric.\\
In the weighing mode, the coil stationary and we assume that the spatial part of the energy is expressed by the terms $\xi(\theta)$ and $\eta(\theta)$, since these terms represent the coupling effect between the electrical elements, and such a coupling is dependent on the relative position of the elements involved. We can assume, for both, a sinusoidal dependence from the angular position $\theta$, such that $\sin\theta\approx \theta$. Finally, the force in the fixed frame is calculated as:
\begin{align}
f_{\theta}=i^Ai^b(\xi+\eta i^A)\label{eq:non_lin_f}
\end{align}
Notice that, when $\eta=0$, we recover the linear equation for a coil immersed in a magnetic field:
\begin{align}
f_{\theta}=i^Ai^b\xi:= \varphi_b i^b:= B\cdot A i^b  \label{eq:lin_f}
\end{align}
where $A$ is the area of the coil.\\
While the derivation of the proposed equations may not be intuitive, the obtained governing equations lead to known results (like the current effects explained in detail in \cite{li2017coil}). However, such equations can be readily manipulated to introduce additional non-linear effects. For instance, mechanical non-linear problems can be studied in the exact same way: the reader will  find out, starting from the proposed equations, how mechanical hysteresis interfere with both force and voltage equations. \\
As a final remark, since all the calculations started from a postulated energy representation, one may object that it would be fairly difficult to correctly measure the energy of the system, or even to give a  representation of the energy itself. Such a question will be answered in an additional paper, where a parallelism with continuum mechanics will be made regarding the introduction of the Helmholtz-like free potential energy which reflects the properties of \emph{material objectivity}. Such a theory proved to be extremely satisfactory in the field of continuum mechanics, which is based on fairly similar mathematical tools as one presented in this paper (see, for instance, \cite{maugin2013continuum} for the theory behind the deformation of electromagnetic solid).

\subsection{Jiles-Aetherton model for hysteresis curves}
The effect of an external current on a permanent magnet causes the shift (more or less evident) of its \emph{working point}. Indeed, demagnetizing effects are serious challenges electro-mechanical machines have to face to ensure proper function of the system and the Kibble Balance does not make an exception, especially for the low uncertainties involved. Therefore, we simulate the BH curves of the permanent magnet using a Jiles-Aetherton model \citep{jiles1986theory} implemented in MATLAB as in \citep{li2019research} for a ferromagnetic material with known magnetic properties. To effectively visualize the difference between linear and non-linear cases, we choose \emph{ad-hoc} parameters (which very rarely may occur in experiments), which yield a highly non-linear ferromagnet. The input current is set to a sinusoidal current:
\begin{align}
i^{b}=I\sin(\omega t)
\end{align}
with frequency $\omega=1\mathrm{Hz}$ (even though, for the purpose model, the frequency does not affect the result regarding relative BH path per period; a current reversal over hours will have the same effect of a reverse over a micro-seconds). The remaining parts are geometry-dependent and represent the machine set-up in terms of dimensions and relative position of the electromagnetic elements. For this, we set some ideal values to give a sense of the terms involved. 

The parameters of the Jiles-Aetherton model are the saturation magnetization $M_s= 1.6 \times 10^6$ A/m, the shape parameter $a=1400$ A/m, the Weiss correction factor $\alpha_{JA}=0.002$ (coupling between the B and H field), the reversibility coefficient $c=0.35$. Other parameters set for the simulations are: 
$\alpha=0.3$ H, $\beta=0.1$ H, $\xi=0.4$ H, $\eta=0.01$ H/A, $I=1.5\div 2$ A.
\\ From the BH curves, the fields are brought to their current equivalents, and the model is applied. The resulting variation of the magnetizing current $i^A$ over the coil-current $i^b$ is shown in figure \ref{fig:ib_ia}: it can be seen how the choice of the parameter in the Jiles-Aetherton model  results in the desired highly non-linear permanent magnet, whose magnetizing current easily responds to the external coil-current $i^b$. The reason for such a large $i^A$ variation is due, in particular, to the parameter $c$ and $a$, which, for a strong magnet, are in the order of $c<0.2$ and $a> 2000$ A/m.\\
As a final remark, in the case of fixed electric elements (weighing mode), the voltage equation for the magnetizing current yields:
\begin{align}
\frac{di^A}{dt}=-\frac{1}{\gamma_{AA}}\bigg(\gamma_{0A}+\gamma_{Ab}i^b+\gamma_{Ab}\frac{di^b}{dt}\bigg)
\end{align}
It can be noticed that the term $\gamma_{AA}$ represents how the PM opposes variations of its magnetizing current. Therefore, the initial interpretation of inductance terms as \emph{electrical} inertia is again shown. In the limiting case of $\gamma_{AA}\to \infty$, the PM is a perfect magnet and $\gamma_{0i}=\gamma_{00}\to 0$ and we recover the linear case.\\
Finally, we see that the induced voltage changes with $i^{A}$; therefore, if the current is changed from $i^b(t_0)=0$ up to $i^{b}_{\mathrm{max}}$ and back to $i^b(t_f)=0$, then $i^{A}(i^{b}(t_f))\neq i^{A}(i^{b}(t_0))$. This consideration is a weakening troublesome point of the Kibble principle.

\subsection{Force variation in weighing mode}
\begin{figure}
\centering
\includegraphics[width=8cm]{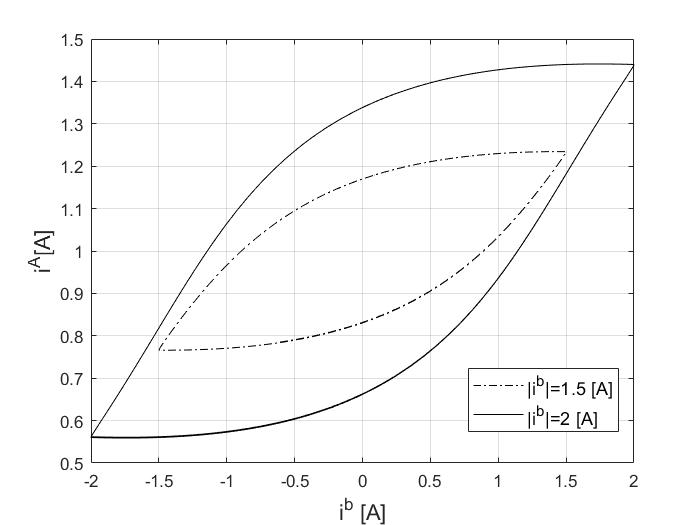}\\
\caption{Magnetizing current $i^A$ over coil-current $i^b$ for $i^b=1.5$ A, $i^b=2$ A, $i^A(0)=1$ A.}\label{fig:ib_ia}
\end{figure}

\begin{figure}
\centering
\includegraphics[width=9cm]{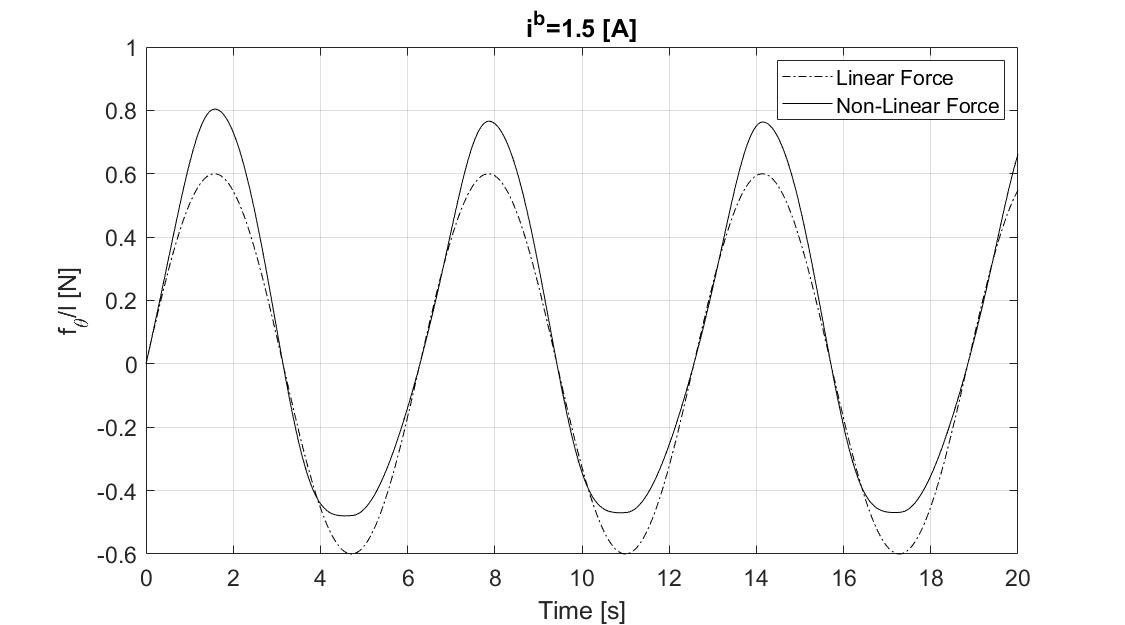}\\
\includegraphics[width=9cm]{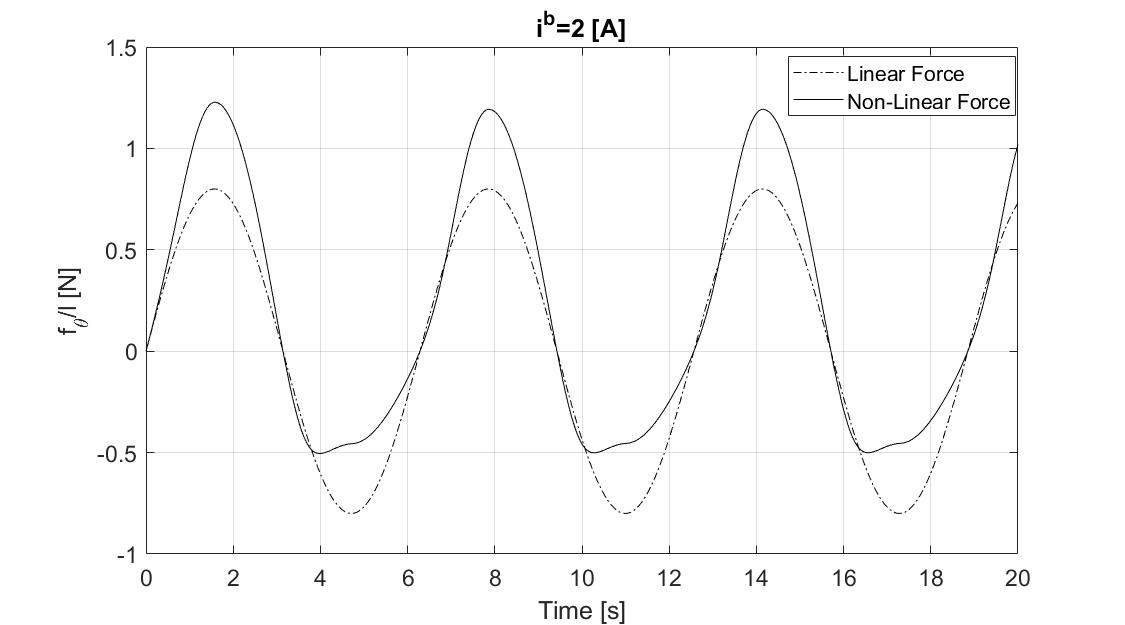}
\caption{Force and non-linear force over time}\label{fig:force_time}
\end{figure}

\begin{figure}
\centering
\includegraphics[width=9cm]{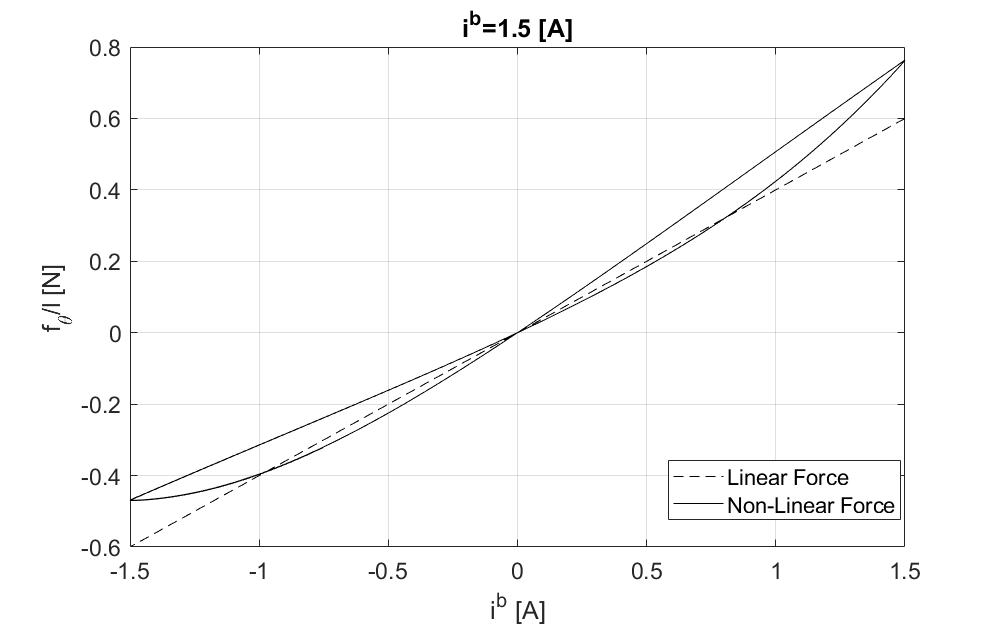}
\includegraphics[width=9cm]{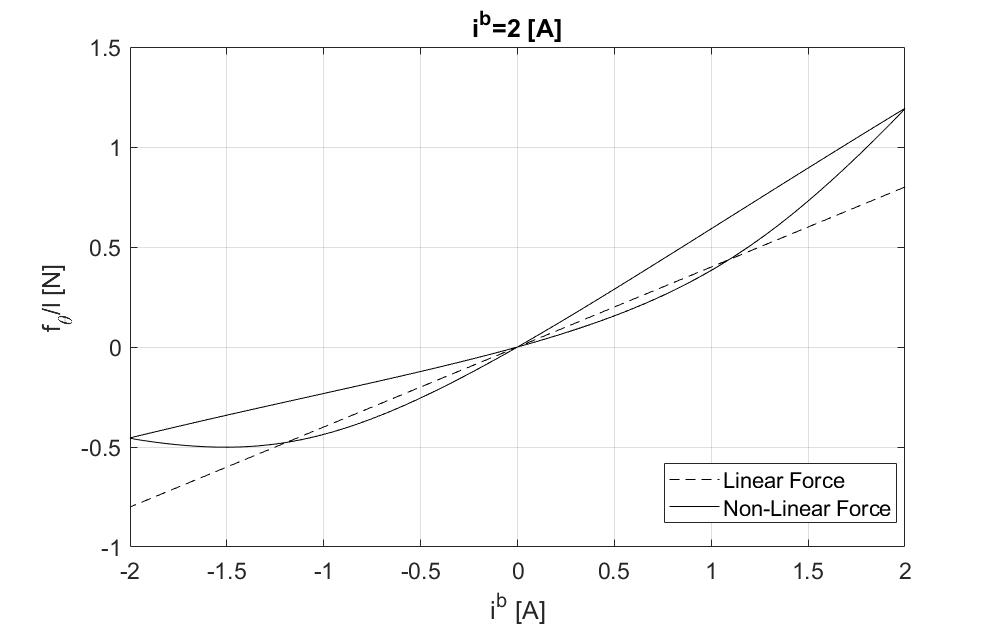}
\caption{Force and non-linear force over coil-current}\label{fig:force_current}
\end{figure}

From \eqref{eq:non_lin_f} and \eqref{eq:lin_f}, we evaluate the torque per unit length $f_{\theta}/l$ which is a force in $[N]$ when the current is changing in a sinusoidal way: this is to simulate the current variation from zero to its maximum value and back to zero, but also the effect of the current reversal.
The comparison between the linear and non-linear force is depicted in figures \ref{fig:force_time}-\ref{fig:force_current}. It can be seen that the coefficient $\eta$ is the one responsible for the non-linear behavior. The non-linear effects increase with the coil current when $\xi$ and $\eta$ are fixed: in this case, far from saturation, the BH curve is highly non-linear and hysteretic: this behavior corresponds to a greater variation in the magnetizing current of the PM (fig.\ref{fig:ib_ia}). Since, for the purpose of illustration, we set the parameter of the Jiles-Aetherton model to yield a highly non-linear ferromagnetic material, we notice that the force reflects the non-linear and hysteretic characteristic of the PM. In particular, the non-linear behavior is characterized by the broken symmetry of the force when the current is reversed; the presence of closed loops expressed the hysteretic behavior (loops pinched correctly at $i^b=0$ A, when the force is zero).\\
As a final note, we remark that in the simulations performed, no dynamical friction nor resistance terms have been taken into account, hence a dissipative behavior can be easily implemented in future analysis accounting for the terms $R_{\alpha\beta}$. 

\subsection{Variation in velocity mode}
From the ad hoc highly-nonlinear simulation, we can see how the the current in the coil affects the permanent magnet in such a way that, once the weighing mode ends, the permanent magnet used for the velocity mode is different from how it was before, since its \emph{working point} shifted to a new state. In this case, it does not make sense to equate the values from the weighing and velocity mode, since they belong, for all means and purposes, to two distinct systems. In conclusion, one can observe that the equation 3 of the paper \cite{li2017coil} and the related corrections of the \emph{coil current effect} can be derived and introduced in the governing equations directly from the mathematical tools presented in this paper.
\section{Conclusion}
In this paper, modern differential geometry is applied to describe non-linear electro-mechanical systems. In particular, we re-interpret Kron's formalism of electric machines with the aid of Finsler geometry. The role of an electro-mechanical metric is introduced to systematically treat general non-linear electro-mechanical systems, thus representing the building blocks of this formalism.
The theoretical apparatus is then applied to the analysis of Kibble Balance systems. Due to the intrinsic non-linear behaviour of these devices, the authors point out that the virtual work principle has to consider the irreversibility constraints to correctly evaluate the design of Kibble Balance machines (refer, for instance, to \cite{inbook, biot1975virtual} for an overview of \emph{virtual dissipation} and Gauss' principle). The authors believe that some of theoretical simplifications employed up to now (and thus the consequent technical rationals coming from the nowadays established theory) represent a significant obstacle to the objective of kilogram (or Newton) redefinition. Therefore, the authors present a general model to rigorously include electrical and mechanical non-linear behaviors. Such a model can be customized to face different non-linear scenarios. Numerical simulations are performed to study the problem of hysteresis effects due to the presence of a time-changing current in the proximity of a permanent magnet. The evolution of linear and non-linear forces are compared, and significant differences appear and are quantified in the case of highly non-linear systems (high current values and weak permanent magnets). As a consequence, we show that the equivalence between velocity mode and weighing mode as a foundation of the working principle of the Kibble balance needs to be refined to systematically account for non-linear terms, as proposed in this paper. The Finsler model proposed can be developed further to take into account additional non-linear behaviors, e.g. mechanical hysteresis and general dissipations. These modifications will be the subject of an additional paper.   
\section*{Conflict of interest}
The Authors declare that they do not have any conflict of interest.

\appendix
\section*{Appendix}
In this section, we will briefly introduce the standard way of deriving the governing equations for a system starting from its Lagrangian and underlying the introduction of the metric tensor (which is usually considered as a unitary diagonal matrix, and therefore neglected).
\subsection{Riemannian metric}
\label{par:riem}
Consider a $N$-Riemannian manifold with the local coordinate $\{z^i\}_{i=1}^{N}$.
The Lagrangian of the system is defined as:
\begin{align}
\mathcal{L}=T-V
\end{align}
where $T$ is the kinetic energy, while $V$ is the potential energy.
If the potential is dependent only on the generalized coordinates, then the Lagrangian is a quadratic form:
\begin{align}
\mathcal{L}=\frac{1}{2}g_{ij}\dot{z}^i\dot{z}^j
\end{align}
since the metric is not velocity/current dependent. The least action principle:
\begin{align}
\delta\int dt \mathcal{L}=0
\end{align}
yields the Euler-Lagrange equations (E-L)
\begin{align}
\frac{d}{dt}\frac{\partial\mathcal{L}}{\partial \dot{z}^{\alpha}}-\frac{\partial \mathcal{L}}{\partial z^{\alpha}}=0
\end{align}
Explicitly, inserting external forces and dissipation terms:
\begin{align}
\frac{d}{dt}(g_{kl}\dot{z}^{l})+r_{kl}\dot{z}^{l}=f_k
\end{align}
where $f^{k}$ represents the sum of integrable and external forces and $r_{kl}$ is the dissipation tensor (resistance and friction tensors). The first term is:
\begin{align}
\frac{d}{dt}(g_{kl}\dot{z}^{l})=\frac{d g_{kl}}{dt}\dot{z}^{l}+g_{kl}\ddot{z}^{l}
\end{align}
If the metric tensor has an implicit time dependence through the generalized coordinate $z^{k}$, then:
\begin{align}
\frac{dg_{kl}}{dt}=\frac{\partial g_{kl}}{\partial z^{\alpha}}\dot{z}^{\alpha}
\end{align}
As in differential geometry, we define the Christoffel symbols:
\begin{align}
\Gamma_{kji}=\frac{1}{2}\bigg(\frac{\partial g_{ji}}{\partial z^{k}}+\frac{\partial g_{ki}}{\partial z^{j}}-\frac{\partial g_{kj}}{\partial z^{i}}\bigg)
\end{align}
Hence, the equations of motion are:
\begin{align}
g_{ij}\ddot{z}^j+\Gamma_{kji}\dot{z}^k\dot{z}^j+r_{ij}\dot{z}^j=f_i\label{eq:riemann}
\end{align}

\def\bibsection{\section*{\refname}} 

\bibliography{apssamp}

\begin{thebibliography}{25}%
\makeatletter
\providecommand \@ifxundefined [1]{%
 \@ifx{#1\undefined}
}%
\providecommand \@ifnum [1]{%
 \ifnum #1\expandafter \@firstoftwo
 \else \expandafter \@secondoftwo
 \fi
}%
\providecommand \@ifx [1]{%
 \ifx #1\expandafter \@firstoftwo
 \else \expandafter \@secondoftwo
 \fi
}%
\providecommand \natexlab [1]{#1}%
\providecommand \enquote  [1]{``#1''}%
\providecommand \bibnamefont  [1]{#1}%
\providecommand \bibfnamefont [1]{#1}%
\providecommand \citenamefont [1]{#1}%
\providecommand \href@noop [0]{\@secondoftwo}%
\providecommand \href [0]{\begingroup \@sanitize@url \@href}%
\providecommand \@href[1]{\@@startlink{#1}\@@href}%
\providecommand \@@href[1]{\endgroup#1\@@endlink}%
\providecommand \@sanitize@url [0]{\catcode `\\12\catcode `\$12\catcode `\&12\catcode `\#12\catcode `\^12\catcode `\_12\catcode `\%12\relax}%
\providecommand \@@startlink[1]{}%
\providecommand \@@endlink[0]{}%
\providecommand \url  [0]{\begingroup\@sanitize@url \@url }%
\providecommand \@url [1]{\endgroup\@href {#1}{\urlprefix }}%
\providecommand \urlprefix  [0]{URL }%
\providecommand \Eprint [0]{\href }%
\providecommand \doibase [0]{https://doi.org/}%
\providecommand \selectlanguage [0]{\@gobble}%
\providecommand \bibinfo  [0]{\@secondoftwo}%
\providecommand \bibfield  [0]{\@secondoftwo}%
\providecommand \translation [1]{[#1]}%
\providecommand \BibitemOpen [0]{}%
\providecommand \bibitemStop [0]{}%
\providecommand \bibitemNoStop [0]{.\EOS\space}%
\providecommand \EOS [0]{\spacefactor3000\relax}%
\providecommand \BibitemShut  [1]{\csname bibitem#1\endcsname}%
\let\auto@bib@innerbib\@empty
\bibitem [{\citenamefont {Mills}\ \emph {et~al.}(2005)\citenamefont {Mills}, \citenamefont {Mohr}, \citenamefont {Quinn}, \citenamefont {Taylor},\ and\ \citenamefont {Williams}}]{mills2005redefinition}%
  \BibitemOpen
  \bibfield  {author} {\bibinfo {author} {\bibfnamefont {I.~M.}\ \bibnamefont {Mills}}, \bibinfo {author} {\bibfnamefont {P.~J.}\ \bibnamefont {Mohr}}, \bibinfo {author} {\bibfnamefont {T.~J.}\ \bibnamefont {Quinn}}, \bibinfo {author} {\bibfnamefont {B.~N.}\ \bibnamefont {Taylor}},\ and\ \bibinfo {author} {\bibfnamefont {E.~R.}\ \bibnamefont {Williams}},\ }\bibfield  {title} {\bibinfo {title} {Redefinition of the kilogram: a decision whose time has come},\ }\href@noop {} {\bibfield  {journal} {\bibinfo  {journal} {Metrologia}\ }\textbf {\bibinfo {volume} {42}},\ \bibinfo {pages} {71} (\bibinfo {year} {2005})}\BibitemShut {NoStop}%
\bibitem [{\citenamefont {Schlamminger}\ \emph {et~al.}(2018)\citenamefont {Schlamminger} \emph {et~al.}}]{schlamminger2018redefining}%
  \BibitemOpen
  \bibfield  {author} {\bibinfo {author} {\bibfnamefont {S.}~\bibnamefont {Schlamminger}} \emph {et~al.},\ }\href@noop {} {\emph {\bibinfo {title} {Redefining the kilogram and other SI units}}}\ (\bibinfo  {publisher} {IOP Publishing},\ \bibinfo {year} {2018})\BibitemShut {NoStop}%
\bibitem [{\citenamefont {Milton}\ \emph {et~al.}(2010)\citenamefont {Milton}, \citenamefont {Williams},\ and\ \citenamefont {Forbes}}]{milton2010quantum}%
  \BibitemOpen
  \bibfield  {author} {\bibinfo {author} {\bibfnamefont {M.~J.}\ \bibnamefont {Milton}}, \bibinfo {author} {\bibfnamefont {J.~M.}\ \bibnamefont {Williams}},\ and\ \bibinfo {author} {\bibfnamefont {A.~B.}\ \bibnamefont {Forbes}},\ }\bibfield  {title} {\bibinfo {title} {The quantum metrology triangle and the redefinition of the si ampere and kilogram; analysis of a reduced set of observational equations},\ }\href@noop {} {\bibfield  {journal} {\bibinfo  {journal} {Metrologia}\ }\textbf {\bibinfo {volume} {47}},\ \bibinfo {pages} {279} (\bibinfo {year} {2010})}\BibitemShut {NoStop}%
\bibitem [{\citenamefont {Kibble}(1976)}]{kibble1976measurement}%
  \BibitemOpen
  \bibfield  {author} {\bibinfo {author} {\bibfnamefont {B.}~\bibnamefont {Kibble}},\ }\bibfield  {title} {\bibinfo {title} {A measurement of the gyromagnetic ratio of the proton by the strong field method},\ }in\ \href@noop {} {\emph {\bibinfo {booktitle} {Atomic masses and fundamental constants 5}}}\ (\bibinfo  {publisher} {Springer},\ \bibinfo {year} {1976})\ pp.\ \bibinfo {pages} {545--551}\BibitemShut {NoStop}%
\bibitem [{\citenamefont {Li}\ \emph {et~al.}(2017)\citenamefont {Li}, \citenamefont {Bielsa}, \citenamefont {Stock}, \citenamefont {Kiss},\ and\ \citenamefont {Fang}}]{li2017coil}%
  \BibitemOpen
  \bibfield  {author} {\bibinfo {author} {\bibfnamefont {S.}~\bibnamefont {Li}}, \bibinfo {author} {\bibfnamefont {F.}~\bibnamefont {Bielsa}}, \bibinfo {author} {\bibfnamefont {M.}~\bibnamefont {Stock}}, \bibinfo {author} {\bibfnamefont {A.}~\bibnamefont {Kiss}},\ and\ \bibinfo {author} {\bibfnamefont {H.}~\bibnamefont {Fang}},\ }\bibfield  {title} {\bibinfo {title} {Coil-current effect in kibble balances: analysis, measurement, and optimization},\ }\href@noop {} {\bibfield  {journal} {\bibinfo  {journal} {Metrologia}\ }\textbf {\bibinfo {volume} {55}},\ \bibinfo {pages} {75} (\bibinfo {year} {2017})}\BibitemShut {NoStop}%
\bibitem [{\citenamefont {Li}\ \emph {et~al.}(2019)\citenamefont {Li}, \citenamefont {Bielsa}, \citenamefont {Stock}, \citenamefont {Kiss},\ and\ \citenamefont {Fang}}]{li2019investigation}%
  \BibitemOpen
  \bibfield  {author} {\bibinfo {author} {\bibfnamefont {S.}~\bibnamefont {Li}}, \bibinfo {author} {\bibfnamefont {F.}~\bibnamefont {Bielsa}}, \bibinfo {author} {\bibfnamefont {M.}~\bibnamefont {Stock}}, \bibinfo {author} {\bibfnamefont {A.}~\bibnamefont {Kiss}},\ and\ \bibinfo {author} {\bibfnamefont {H.}~\bibnamefont {Fang}},\ }\bibfield  {title} {\bibinfo {title} {An investigation of magnetic hysteresis error in kibble balances},\ }\href@noop {} {\bibfield  {journal} {\bibinfo  {journal} {IEEE Transactions on Instrumentation and Measurement}\ }\textbf {\bibinfo {volume} {69}},\ \bibinfo {pages} {5717} (\bibinfo {year} {2019})}\BibitemShut {NoStop}%
\bibitem [{\citenamefont {Keck}\ \emph {et~al.}(2022)\citenamefont {Keck}, \citenamefont {Seifert}, \citenamefont {Newell}, \citenamefont {Schlamminger}, \citenamefont {Theska},\ and\ \citenamefont {Haddad}}]{keck2022design}%
  \BibitemOpen
  \bibfield  {author} {\bibinfo {author} {\bibfnamefont {L.}~\bibnamefont {Keck}}, \bibinfo {author} {\bibfnamefont {F.}~\bibnamefont {Seifert}}, \bibinfo {author} {\bibfnamefont {D.}~\bibnamefont {Newell}}, \bibinfo {author} {\bibfnamefont {S.}~\bibnamefont {Schlamminger}}, \bibinfo {author} {\bibfnamefont {R.}~\bibnamefont {Theska}},\ and\ \bibinfo {author} {\bibfnamefont {D.}~\bibnamefont {Haddad}},\ }\bibfield  {title} {\bibinfo {title} {Design of an enhanced mechanism for a new kibble balance directly traceable to the quantum si},\ }\href@noop {} {\bibfield  {journal} {\bibinfo  {journal} {EPJ Techniques and Instrumentation}\ }\textbf {\bibinfo {volume} {9}},\ \bibinfo {pages} {7} (\bibinfo {year} {2022})}\BibitemShut {NoStop}%
\bibitem [{\citenamefont {Kron}(1934)}]{kron1934non}%
  \BibitemOpen
  \bibfield  {author} {\bibinfo {author} {\bibfnamefont {G.}~\bibnamefont {Kron}},\ }\bibfield  {title} {\bibinfo {title} {Non-riemannian dynamics of rotating electrical machinery},\ }\href@noop {} {\bibfield  {journal} {\bibinfo  {journal} {Journal of mathematics and physics}\ }\textbf {\bibinfo {volume} {13}},\ \bibinfo {pages} {103} (\bibinfo {year} {1934})}\BibitemShut {NoStop}%
\bibitem [{\citenamefont {Lynn}(1958)}]{lynn1958tensor}%
  \BibitemOpen
  \bibfield  {author} {\bibinfo {author} {\bibfnamefont {J.}~\bibnamefont {Lynn}},\ }\bibfield  {title} {\bibinfo {title} {Tensor analysis of electrical machine hunting},\ }\href@noop {} {\bibfield  {journal} {\bibinfo  {journal} {Proceedings of the IEE-Part C: Monographs}\ }\textbf {\bibinfo {volume} {105}},\ \bibinfo {pages} {420} (\bibinfo {year} {1958})}\BibitemShut {NoStop}%
\bibitem [{\citenamefont {Gibbs}(1952)}]{gibbs1952tensors}%
  \BibitemOpen
  \bibfield  {author} {\bibinfo {author} {\bibfnamefont {W.~J.}\ \bibnamefont {Gibbs}},\ }\href@noop {} {\emph {\bibinfo {title} {Tensors in electrical machine theory}}}\ (\bibinfo  {publisher} {Chapman \& Hall},\ \bibinfo {year} {1952})\BibitemShut {NoStop}%
\bibitem [{\citenamefont {Gibbs}(1965)}]{gibbs1965foundation}%
  \BibitemOpen
  \bibfield  {author} {\bibinfo {author} {\bibfnamefont {W.}~\bibnamefont {Gibbs}},\ }\bibfield  {title} {\bibinfo {title} {Foundation of modern network theories},\ }\href@noop {} {\bibfield  {journal} {\bibinfo  {journal} {Nature}\ }\textbf {\bibinfo {volume} {207}},\ \bibinfo {pages} {1118} (\bibinfo {year} {1965})}\BibitemShut {NoStop}%
\bibitem [{\citenamefont {Von Der~Embse}(1968)}]{von1968new}%
  \BibitemOpen
  \bibfield  {author} {\bibinfo {author} {\bibfnamefont {U.~A.}\ \bibnamefont {Von Der~Embse}},\ }\bibfield  {title} {\bibinfo {title} {New theory of nonlinear commutator machines},\ }\href@noop {} {\bibfield  {journal} {\bibinfo  {journal} {IEEE Transactions on Power Apparatus and Systems}\ ,\ \bibinfo {pages} {1804}} (\bibinfo {year} {1968})}\BibitemShut {NoStop}%
\bibitem [{\citenamefont {Hamaji}\ \emph {et~al.}(2021)\citenamefont {Hamaji}, \citenamefont {Nishino},\ and\ \citenamefont {Ogushi}}]{hamaji2021design}%
  \BibitemOpen
  \bibfield  {author} {\bibinfo {author} {\bibfnamefont {M.}~\bibnamefont {Hamaji}}, \bibinfo {author} {\bibfnamefont {A.}~\bibnamefont {Nishino}},\ and\ \bibinfo {author} {\bibfnamefont {K.}~\bibnamefont {Ogushi}},\ }\bibfield  {title} {\bibinfo {title} {Design of a new dynamic torque generation machine based on the principle of kibble balance},\ }\href@noop {} {\bibfield  {journal} {\bibinfo  {journal} {Measurement: Sensors}\ }\textbf {\bibinfo {volume} {18}},\ \bibinfo {pages} {100183} (\bibinfo {year} {2021})}\BibitemShut {NoStop}%
\bibitem [{\citenamefont {Abraham}\ and\ \citenamefont {Marsden}(2008)}]{abraham2008foundations}%
  \BibitemOpen
  \bibfield  {author} {\bibinfo {author} {\bibfnamefont {R.}~\bibnamefont {Abraham}}\ and\ \bibinfo {author} {\bibfnamefont {J.~E.}\ \bibnamefont {Marsden}},\ }\href@noop {} {\emph {\bibinfo {title} {Foundations of mechanics}}},\ \bibinfo {number} {364}\ (\bibinfo  {publisher} {American Mathematical Soc.},\ \bibinfo {year} {2008})\BibitemShut {NoStop}%
\bibitem [{\citenamefont {Bossavit}(1988)}]{bossavit1988whitney}%
  \BibitemOpen
  \bibfield  {author} {\bibinfo {author} {\bibfnamefont {A.}~\bibnamefont {Bossavit}},\ }\bibfield  {title} {\bibinfo {title} {Whitney forms: A class of finite elements for three-dimensional computations in electromagnetism},\ }\href@noop {} {\bibfield  {journal} {\bibinfo  {journal} {IEE Proceedings A (Physical Science, Measurement and Instrumentation, Management and Education, Reviews)}\ }\textbf {\bibinfo {volume} {135}},\ \bibinfo {pages} {493} (\bibinfo {year} {1988})}\BibitemShut {NoStop}%
\bibitem [{\citenamefont {Kond{\=o}}(1955)}]{kondo1955raag}%
  \BibitemOpen
  \bibfield  {author} {\bibinfo {author} {\bibfnamefont {K.}~\bibnamefont {Kond{\=o}}},\ }\href@noop {} {\emph {\bibinfo {title} {RAAG memoirs of the unifying study of basic problems in engineering and physical sciences by means of geometry}}},\ Vol.~\bibinfo {volume} {3}\ (\bibinfo  {publisher} {Gakujutsu Bunken Fukyu-kai.},\ \bibinfo {year} {1955})\BibitemShut {NoStop}%
\bibitem [{\citenamefont {Ratliff}\ \emph {et~al.}(2021)\citenamefont {Ratliff}, \citenamefont {Van~Wyk}, \citenamefont {Xie}, \citenamefont {Li},\ and\ \citenamefont {Rana}}]{ratliff2021generalized}%
  \BibitemOpen
  \bibfield  {author} {\bibinfo {author} {\bibfnamefont {N.~D.}\ \bibnamefont {Ratliff}}, \bibinfo {author} {\bibfnamefont {K.}~\bibnamefont {Van~Wyk}}, \bibinfo {author} {\bibfnamefont {M.}~\bibnamefont {Xie}}, \bibinfo {author} {\bibfnamefont {A.}~\bibnamefont {Li}},\ and\ \bibinfo {author} {\bibfnamefont {M.~A.}\ \bibnamefont {Rana}},\ }\bibfield  {title} {\bibinfo {title} {Generalized nonlinear and finsler geometry for robotics},\ }in\ \href@noop {} {\emph {\bibinfo {booktitle} {2021 IEEE International Conference on Robotics and Automation (ICRA)}}}\ (\bibinfo {organization} {IEEE},\ \bibinfo {year} {2021})\ pp.\ \bibinfo {pages} {10206--10212}\BibitemShut {NoStop}%
\bibitem [{\citenamefont {Boccaletti}\ \emph {et~al.}(1997)\citenamefont {Boccaletti}, \citenamefont {Di~Bari}, \citenamefont {Cipriani},\ and\ \citenamefont {Pucacco}}]{boccaletti1997finsler}%
  \BibitemOpen
  \bibfield  {author} {\bibinfo {author} {\bibfnamefont {D.}~\bibnamefont {Boccaletti}}, \bibinfo {author} {\bibfnamefont {M.}~\bibnamefont {Di~Bari}}, \bibinfo {author} {\bibfnamefont {P.}~\bibnamefont {Cipriani}},\ and\ \bibinfo {author} {\bibfnamefont {G.}~\bibnamefont {Pucacco}},\ }\bibfield  {title} {\bibinfo {title} {Finsler geometry in classical mechanics and in bianchi cosmological models.},\ }\href@noop {} {\bibfield  {journal} {\bibinfo  {journal} {Nuovo Cimento B Serie}\ }\textbf {\bibinfo {volume} {112}},\ \bibinfo {pages} {213} (\bibinfo {year} {1997})}\BibitemShut {NoStop}%
\bibitem [{\citenamefont {Kawaguchi}(1977)}]{kawaguchi1977application}%
  \BibitemOpen
  \bibfield  {author} {\bibinfo {author} {\bibfnamefont {T.}~\bibnamefont {Kawaguchi}},\ }\bibfield  {title} {\bibinfo {title} {On the application of finsler geometry to engineering dynamical systems},\ }\href@noop {} {\bibfield  {journal} {\bibinfo  {journal} {Periodica Mathematica Hungarica}\ }\textbf {\bibinfo {volume} {8}},\ \bibinfo {pages} {281} (\bibinfo {year} {1977})}\BibitemShut {NoStop}%
\bibitem [{\citenamefont {Li}\ and\ \citenamefont {Schlamminger}(2022)}]{li2022irony}%
  \BibitemOpen
  \bibfield  {author} {\bibinfo {author} {\bibfnamefont {S.}~\bibnamefont {Li}}\ and\ \bibinfo {author} {\bibfnamefont {S.}~\bibnamefont {Schlamminger}},\ }\bibfield  {title} {\bibinfo {title} {The irony of the magnet system for kibble balances—a review},\ }\href@noop {} {\bibfield  {journal} {\bibinfo  {journal} {Metrologia}\ }\textbf {\bibinfo {volume} {59}},\ \bibinfo {pages} {022001} (\bibinfo {year} {2022})}\BibitemShut {NoStop}%
\bibitem [{\citenamefont {Maugin}(2013)}]{maugin2013continuum}%
  \BibitemOpen
  \bibfield  {author} {\bibinfo {author} {\bibfnamefont {G.~A.}\ \bibnamefont {Maugin}},\ }\href@noop {} {\emph {\bibinfo {title} {Continuum mechanics of electromagnetic solids}}}\ (\bibinfo  {publisher} {Elsevier},\ \bibinfo {year} {2013})\BibitemShut {NoStop}%
\bibitem [{\citenamefont {Jiles}\ and\ \citenamefont {Atherton}(1986)}]{jiles1986theory}%
  \BibitemOpen
  \bibfield  {author} {\bibinfo {author} {\bibfnamefont {D.~C.}\ \bibnamefont {Jiles}}\ and\ \bibinfo {author} {\bibfnamefont {D.~L.}\ \bibnamefont {Atherton}},\ }\bibfield  {title} {\bibinfo {title} {Theory of ferromagnetic hysteresis},\ }\href@noop {} {\bibfield  {journal} {\bibinfo  {journal} {Journal of magnetism and magnetic materials}\ }\textbf {\bibinfo {volume} {61}},\ \bibinfo {pages} {48} (\bibinfo {year} {1986})}\BibitemShut {NoStop}%
\bibitem [{\citenamefont {Li}\ and\ \citenamefont {Gong}(2019)}]{li2019research}%
  \BibitemOpen
  \bibfield  {author} {\bibinfo {author} {\bibfnamefont {Z.}~\bibnamefont {Li}}\ and\ \bibinfo {author} {\bibfnamefont {Y.}~\bibnamefont {Gong}},\ }\bibfield  {title} {\bibinfo {title} {Research on ferromagnetic hysteresis of a magnetorheological fluid damper},\ }\href@noop {} {\bibfield  {journal} {\bibinfo  {journal} {Frontiers in Materials}\ }\textbf {\bibinfo {volume} {6}},\ \bibinfo {pages} {111} (\bibinfo {year} {2019})}\BibitemShut {NoStop}%
\bibitem [{\citenamefont {Wieber}(2007)}]{inbook}%
  \BibitemOpen
  \bibfield  {author} {\bibinfo {author} {\bibfnamefont {P.}~\bibnamefont {Wieber}},\ }\bibinfo {title} {Holonomy and nonholonomy in the dynamics of articulated motion}\ (\bibinfo {year} {2007})\ pp.\ \bibinfo {pages} {411--425}\BibitemShut {NoStop}%
\bibitem [{\citenamefont {Biot}(1975)}]{biot1975virtual}%
  \BibitemOpen
  \bibfield  {author} {\bibinfo {author} {\bibfnamefont {M.~A.}\ \bibnamefont {Biot}},\ }\bibfield  {title} {\bibinfo {title} {A virtual dissipation principle and lagrangian equations in non-linear irreversible thermodynamics},\ }\href@noop {} {\bibfield  {journal} {\bibinfo  {journal} {Bulletins de l'Acad{\'e}mie Royale de Belgique}\ }\textbf {\bibinfo {volume} {61}},\ \bibinfo {pages} {6} (\bibinfo {year} {1975})}\BibitemShut {NoStop}%
\end{thebibliography}%

\end{document}